\documentclass[%
 reprint,
superscriptaddress,
 amsmath,amssymb,
 aps,
]{revtex4-2}

\usepackage{graphicx}
\usepackage{dcolumn}
\usepackage{bm}
\usepackage{hyperref}
\hypersetup{colorlinks=true, citecolor=blue, urlcolor=blue, linkcolor=blue}

\usepackage[alsoload=hep]{siunitx} 

\usepackage{xcolor} 

\usepackage[caption=false]{subfig}

\begin{document}

\preprint{APS/123-QED}

\title{Mu$\chi$e - A Search for Familons in Muon Decay Using HPGe Detectors}

\author{D.~Koltick}
\email{koltick@purdue.edu} 

\author{S.~Huang}

\author{F.~Bergin}

\author{J.~Chen}

\author{H.~Cao}
\address{Department of Physics and Astronomy, Purdue University, West Lafayette, IN 47907, USA}

\date{\today}

\begin{abstract}

\noindent A broken lepton family symmetry can lead to muon decay to a Familon 
\[\mu  \to {\chi _{Familon}} + e .\] We propose\footnote{Presented at Potential Fermilab Muon Campus and Storage Ring Experiments Workshop, May 24-27, 2021} to search for two body  muon decay to a Familon  in the mass range from 86 to 105 MeV/c$^2$, by observing stopped positive muons. This mass region is the search window blind spot of magnetic spectrometers. The search signal will be a mono-energetic positron peak on top of the standard model muon Michel decay spectrum.  The decay positron spectrum will be measured in HPGe detectors surrounded by large volume NaI detectors and a charged particle tracking system used to help reduce false triggers. This short-term experiment requires a stopped positive muon rate of 200~cps with a trigger to stopped muon ratio of 10:1 or better.  At this rate, in a period of 2-hours previous limits will be matched. With 1-year of data collection muon branching ratio limits will range between 10$^{-6}$ to 10$^{-7}$. This experiment takes advantage of Fermilab's higher energy muons to greatly suppress backgrounds for Familon masses near the muon mass compared to lower energy muon factories.\\

\end{abstract}

\maketitle


\section{\label{sec:over} Introduction to the Familon}

\noindent A major puzzle of the standard model is the observation of quark and lepton replicated families. There are 3 possible options to explain family symmetry [1-4], via; discrete symmetries, continuous and local gauge symmetries, or finally, continuous and global gauge symmetries. It is the possibility of broken global family symmetries which leads to Nambu-Goldstone bosons, called \textbf{\textit {Familons}}. The coupling of Familons at low energy is determined by the non-linear realization of the family symmetry \cite{wilczek,dicus,adler,feng}, 
\[\frac{1}{F}{\partial _\mu }{f^a}\bar \varphi _L^i{\gamma ^\mu }T_{ij}^a\varphi _L^j\]
Where F is the family symmetry breaking scale, f$^a$ are the Familons, and T$^a$ are the generators of the broken symmetry. This broken lepton family symmetry can lead to the two body muon decay
\[\mu  \to {\chi _{Familon}} + e  .\]
\noindent A number of searches have been carried out~\cite{bryman,bilger,derenzo,bayes,aguilar}, based on observing the mono-energetic positron and assume the Familon escapes the detection system.  The best limits have achieved  $\sim$10$^{-5}$ muon branching ratio.  However, these are set by observing stopped muons in magnetic spectrometers. The failing of these instruments is the inability to track low energy positrons having kinetic energy in the sub-MeV to $\sim$10 MeV range. This deficiency leaves a hole in the search window for Familon masses near the mass of the muon.  This region is accessible to scintillation or crystal detectors which both contain the stopped muon and observe its decay products. The most accurate experiment reported in the search window hole was completed in the 1960s using the Chicago bubble chamber \cite{derenzo}.\\
\\
\noindent We propose to search for two body muon decay to a Famion in the mass range window blind spot of magnetic spectrometers from 86 to 105 MeV/c$^2$.  The search signal will be a mono-energetic positron peak on top of the standard model muon Michel decay spectrum. It is assumed that the Familon leaves the crystal without decay. The positron spectrum will be measured in HPGe detectors surrounded by large volume NaI detectors and a charged particle tracking system used to help reduce false triggers. This short-term experiment requires a stopped positive muon rate of 200 cps with a trigger to stopped muon ratio of 10:1 or better. At this rate, in a period of 2-hours previous limits will be matched. With 1-year of data collection muon branching ratio limits will range between 10$^{-6}$ to 10$^{-7}$. This experiment takes advantage of Fermilab's higher energy muons to greatly suppress backgrounds for Familon masses near the muon mass compared to lower energy muon factories.
\section{\label{sec:over} Past Familon Search Experiments}
\begin{figure}[]
\includegraphics[width=0.95\columnwidth]{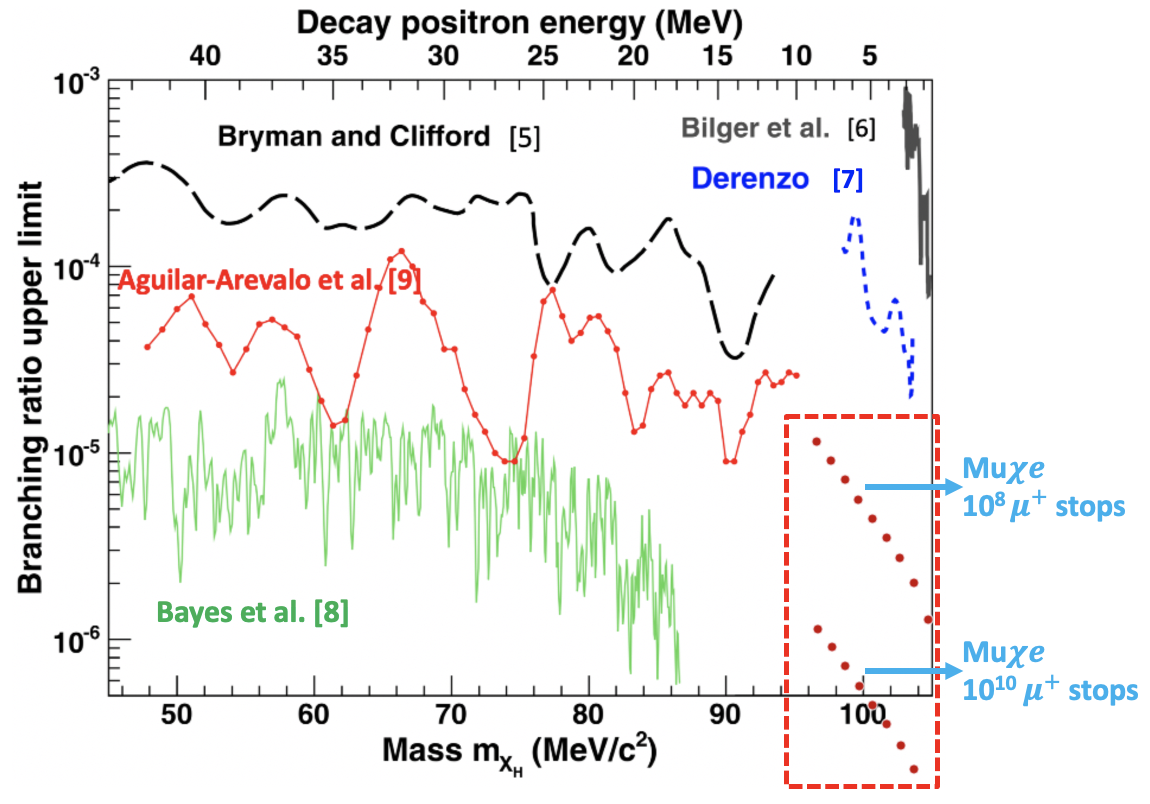}
\caption{\label{fig:90CL}Summary of the experimental 90\% CL branching ratio limits for past experiments.  Insert:  90\% CL branching ratio limits achievable by Mu$\chi$e at Fermilab, assuming 10$^8$ (1 week data) and 10$^{10}$ (1.5 years data) total $\mu^+$ stops at 200 stops/sec.}
\label{fig:limits}
\end{figure}

\begin{figure}[]
\centering
\includegraphics[width=0.25\textwidth]{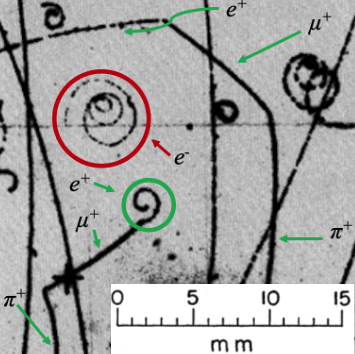}
\caption{\label{derenzo}From the Chicago bubble chamber Familon Search.  Circled in red is a 1.4 MeV/c internal conversion electron emerging from a Mylar strip (faint horizontal line). Highlighted in green are two $\pi^+$-$\mu^+$-e$^+$ decay chains. Circled in green is a 0.7 MeV/c positron. A muon decay to a Familon would have a similar appearance.}
\label{fig:chicago}
\end{figure}

\noindent A number of past experiments, between the years 1969 to 2015 have completed two body muon decay searches to a Familon. Those performed at TRIUMF \cite{bryman,bayes,aguilar} between 1986 and 2015 are based on magnetic spectrometers.  The 90\% confidence sensitivity limits for these experiments are displayed in Figure \ref{fig:limits}. The sensitivity drops to none near M$_\chi\sim$95 MeV/c$^2$ and above.  The reason for this hole in the search window is best illustrated by the 1969 Chicago bubble chamber experiment \cite{derenzo} which measured 2 million bubble chamber tracks.  A low energy positron track is shown in Figure \ref{fig:chicago}.  Such low energy tracks can not pass through the magnetic spectrometer tracking system.  Another approach is to use HPGe detectors which contain the stopped muon and measure the total energy of its decay positron.  A first attempt at this approach was completed at PSI \cite{bilger}.  However, neither of these experiments were able to close the search window hole. \\
\\
\noindent Up coming muon experiments, such as Mu2e and Comet, while planning to reach branching ratio sensitivities at the level of 10$^{-17}$ have no acceptance for muon two body Familon decays. Presently, the Mu2e collaboration is discussing the possibility of searching for Familons with M$_\chi\leq$ 60 MeV/c$^2$ during a special few day calibration run.  But again, such searches do not address the hole in the Familon mass window.

\section{\label{sec:over} M\lowercase{u}$\chi$\lowercase{e} Experimental Layout}
\noindent The Mu$\chi$e experiment will utilize the g-2 $\mu^+$ beam as its muon source. Here we simply summarize the beam production and transportation described in detail in Ref.~\cite{stratakis}. The production of the beam starts from 8.9 GeV protons delivered through the M1 beam line to an Inconel target, producing a spectrum of secondary particles. A downstream pulsed dipole magnet then selects 3.1 GeV/c positive particles, propagating through the M2, M3 line channel and the delivery ring, and finally reaches the M5 beam line area.\\

\noindent The layout of the experiment consists of HPGe detectors surrounded by large volume NaI detectors and a charged particle tracking system used to point the incoming muon through the absorber and energy degrader to a position on the face of the HPGe detection system. \\

\noindent The beam passes through a large initial NaI trigger  detector which demands a high energy, $\sim$100 MeV energy deposition. Such an energy deposition is far above natural and induced radiation backgrounds, assuring the presence of a beam particle. The beam then passes through an absorber to filter any other particle species from the beam. In addition it serves to reduce the muon energy. The absorber is followed by an additional trigger NaI detector, again demanding a $\sim$100 MeV energy deposition.  Charged particle tracking chambers are placed on either side of the absorber to point muon tracks through the absorber and to estimate the stopping location within the HPGe detection system.\\

\noindent The HPGe detectors serve to finally stop the incoming muon and to observe the decay positron.  If cylindrical commercial HPGe detectors are used, they would view the beam sideways not face-on and arranged staggered for more uniform acceptance along the beam axis. The issue using such large individual detectors is the dynamic range between the deposited stopping muon energy, up to $\sim$100 MeV, and the lower $\sim$1 MeV positron kinetic energy of interest.  To reduce the dynamic range a single detector requires, an HPGe array of thin detectors can be used. This allows a maximum muon deposition of $\sim$20 MeV in a single detector. The same effect can be achieved using commercial detectors but with a loss of acceptance.\\

\noindent The HPGe detection system is then surrounded by an array of larger NaI detectors that serve to veto muon decay events not fully contained within the HPGe detector system. Figure \ref{fig:pulse} shows such an event collected from a stopped cosmic ray muon in the Purdue M\lowercase{u}$\chi$\lowercase{e} test stand.  
\begin{figure}[]
\includegraphics[width=0.85\columnwidth]{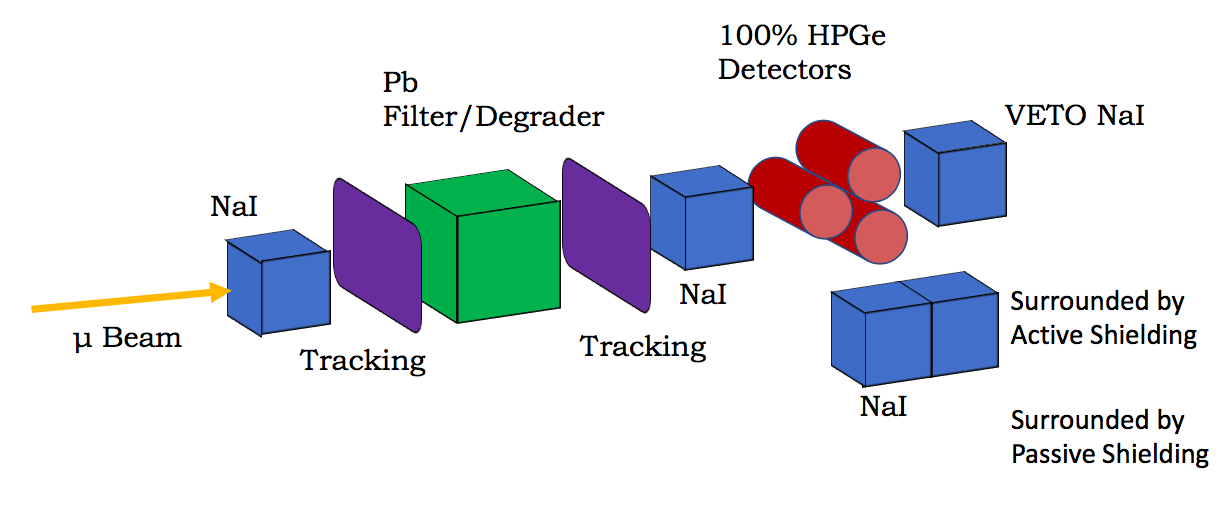}
\caption{\label{fig:layout}Mu$\chi$e schematic layout. The high energy Fermilab beam allows for greatly reduced backgrounds. The HPGe detectors are surrounded by both an active NaI detector system and a passive radiation shielding system. Multiple cylindrical commercial HPGe detectors are shown in this layout.}
\end{figure}

\begin{figure}[]
\centering
\includegraphics[width=0.3\textwidth]{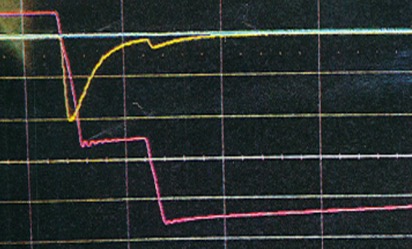}
\caption{\label{fig:pulse}Purdue M\MakeLowercase{u}$\chi$\MakeLowercase{e} test stand event captured by an oscilloscope. Red trace: HPGe detector signal of stopping muon with delayed decay to positron. Yellow trace: NaI trigger counter with in time large muon signal, followed by detection of energy not contained in the HPGe detector creating an event veto. Flat green trace: Additional NaI detector registering no escaping energy.}
\end{figure} 
\section{\label{sec:over}M\lowercase{u}$\chi$\lowercase{e} Event and Search Signal}
\begin{figure}[]
\includegraphics[width=0.85\columnwidth]{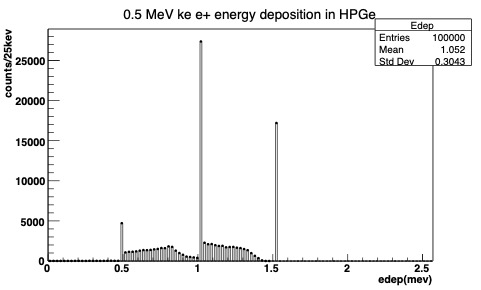}
\caption{\label{fig:sig_finger_print}1-M GEANT4 simulated muon to Familon decays assuming a mono-energetic positron kinetic energy of 0.5 MeV deposited in a 100\% HPGe detector. Three distinct peaks form the fingerprint of a Famlion signal. Shown are the full energy peak (KE+2m$_e$=1.52 MeV), the first escape peak (KE+m$_e$=1.01 MeV) and the second escape peak (KE=0.5 MeV).}
\end{figure}

\noindent The Familon signal produced in the HPGe system is displayed in Figure \ref{fig:sig_finger_print}.  If the complete energy of the decay positron were contained within the HPGe system the Familon would appear as a single line on top of the Michel muon decay spectrum. Because of the high energy depositions required in the trigger counters and the tracking system pointing to the HPGe detector the only background to the signal is the Michel spectrum altered by the energy resolution function of the detectors. However, even with energy loss in the HPGe system, a unique fingerprint of three lines is measured, with fixed ratios, proving that the signal origin is a positron with a fixed kinetic energy.\\
\\
\noindent Key to the search sensitivity is the excellent energy resolution of the HPGe detectors, expected to be 1 keV at 1 MeV or $\delta$E$_\gamma$/E$_\gamma\sim$0.1\%.  Such excellent resolution greatly reduces the background under the signal compared to NaI detectors at $\sim$8\% or even LaBr$_3$ detectors at $\sim$1\%. However, to achieve such resolution requires long pulse integration times, 60 $\mu$sec, causing rate limitations. This coupled with the possible large energy deposition of an overlapping 2$^{nd}$ arriving muon, the electronics dynamic range can be overwhelmed, spoiling the ability to measure precisely the low energy positron at $\sim$1 MeV. In addition the desire to accurately measure the positive muon lifetime,  2.2 $\mu$-sec, out to 5$\tau$ as a check of the experiments systematic errors, again set a speed limit on the muon stopping rate. For these reasons event slots are selected to be $\sim$0.5~msec, so that individual events do not over lap.
\section{\label{sec:over}Test Station Results}
\begin{figure}[]
\includegraphics[width=0.7\columnwidth]{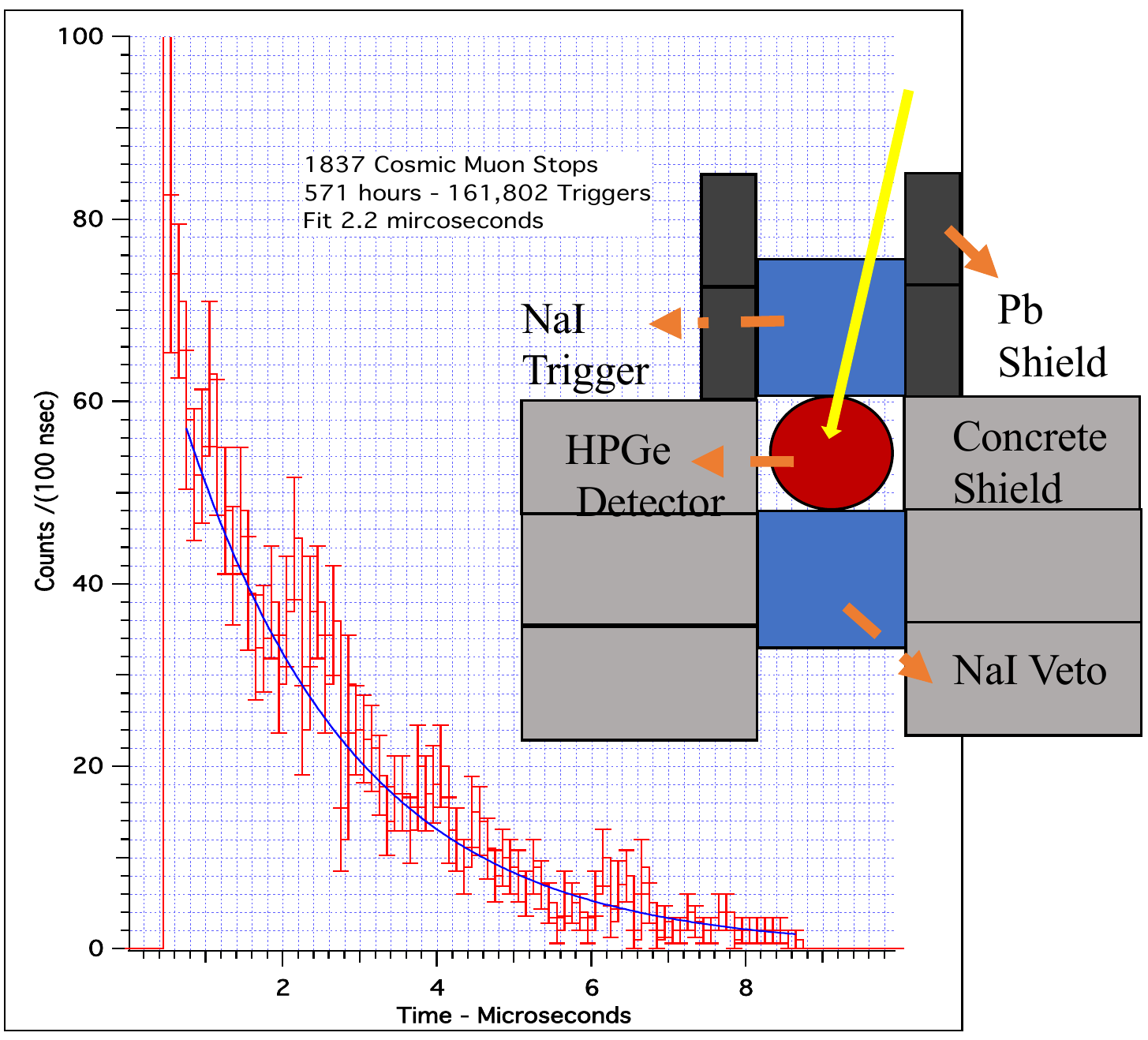}
\caption{\label{fig:cosmic_test}Insert: Purdue M\MakeLowercase{u}$\chi$\MakeLowercase{e} cosmic ray test stand using high energy muons. The 5.5 inch cube NaI detectors serve as both trigger and veto counters. Data collected with 100\% HPGe detector. Graph: Lifetime measurement yielding background to signal for this incomplete system of 10$^{-2}$.   }
\end{figure}

\noindent The search for Familon particles with mass near the muon requires the detection of low energy positrons susceptible to be mimicked by low energy environmental decay background and low energy machine associated background.  To test the rejection advantage that Fermilab allows, a cosmic ray test stand was built based on 5.5 inch cubic NaI detectors. A muon passing through these detectors generates 100 MeV signal far above environmental background radiation and even machine associated background.  The insert in Figure \ref{fig:cosmic_test} show the test stand configuration. This system did not have the full NaI veto system nor a complete passive shielding system yet achieved a signal to background ratio of 10$^{-2}$, found by fitting the lifetime spectrum of events. The background spectrum events also shown in Figure 6 was caused by the large trigger to stop ratio of 100:1. This would be greatly improve by the proposed Fermilab Mu$\chi$e detector shown in Figure \ref{fig:layout} by a factor of 10 to 100. It should be remembered the $\mu^-$ lifetime in Ge is $<$ 200 nsec, so did not contribute to this background estimate. 
\section{\label{sec:over}M\lowercase{u}$\chi$\lowercase{e} Familon Limits}
\noindent
\begin{table}[h]
\caption{\label{tab:limits}%
90\% CL limits over the Familon Mass Range.}
\begin{ruledtabular}
\begin{tabular}{lccc}
\textrm{$\mu$~Stops\footnote{Based on a single 100\% HPGe detector}}&
\textrm{M$_\chi$=95 MeV}&
\multicolumn{1}{c}{\textrm{M$_\chi$=m$_\mu$-m$_e$}}&
\textrm{Time (200 cps)}\\
\colrule
10$^8$ & 2 10$^{-5}$  & 1 10$^{-6}$ & 1 week\\
10$^{10}$ & 2 10$^{-6}$ & 1 10$^{-7}$ & 1.5 years\\
10$^{11}$ & 6 10$^{-7}$ & 3 10$^{-8}$ & (2 kcps) 1.5 year\\
\end{tabular}
\end{ruledtabular}
\end{table}
To  estimate the 90\% confidence level (CL) limit for Mu$\chi$e the Michel decay spectrum for the appropriate number of muon stops was generated. The shape of the spectrum is assumed to be well known so that only statistical error contributes to the limits. The signal expectation was found assuming a search comb with a window given by
\[Window = \left[ {{E_{1st - escape}} \pm 2\sigma } \right] + \left[ {{E_{full - energy}} \pm 2\sigma } \right].\]The double escape peak was dropped from this search due to its low probability. The limit was set when the signal strength in the search window reached 1.3$\sigma_{statistical}$. The limit is based on a uniform and constant muon stopping rate. In addition, because the low singles rate, 2kcps, the HPGe detector is assumed to have no dead-time.  The 100\% HPGe photopeak acceptance and energy leakage were all modeled.
Also, the fake signal rate caused by environmental backgrounds are assumed to be ignorable.\\
\\
\noindent Figure~\ref{fig:90CL} shows the expected 90\% CL limit for 10$^8$ and 10$^{10}$ $\mu$ stops, corresponding to 1 week and 1.5 years of continuous data taking, respectively.  The $\mu$ stop rate is assumed to be 200~cps in a single 100\% HPGe detector.  The limit values at the extremes of the mass window hole obtainable by Mu$\chi$e are displayed in Table \ref{tab:limits}. If a 1:1 trigger to $\mu$ stop ratio can be achieved then 2 kcps of $\mu$ stops would be possible, allowing for 10$^{11}$ stops to be collected in 1.5 years. With this dataset a discovery (5$\sigma$) branching ratio sensitivity of $\Gamma_\chi\sim$ 1$\times$10$^{-7}$ for a Familon having M$_\chi$=m$_\mu$-m$_e$ is within reach.

\begin{acknowledgments}
\noindent
This work has been supported by the Department of Energy contract DE-SC0007884 Intensity Frontier. Additional support for the detectors and electronic readout has been provided by Tech-Source Inc., Germantown, MD 20874 and Rapiscan Systems, Torrance, CA 90503
\end{acknowledgments}

\nocite{*}

\bibliography{PFMCSRE-workshop-template-revtex4-2/PFMCSRE-workshop-template-revtex4-2}

\end{document}